\documentclass[sigconf,preprint]{acmart}

\usepackage{algorithm2e}
\RestyleAlgo{ruled}
\usepackage{amsmath, amsfonts}
\usepackage{bm}
\usepackage{enumitem}
\usepackage{physics}
\usepackage{dirtytalk}

\newcommand{\todo}[1]{}

\newcommand{\DurrHoyer}{D\"{u}rr-H\o{}yer }
\newcommand{\etc}{\emph{etc}.}

\copyrightyear{2025}
\acmYear{2025}

\makeatletter
\renewcommand{\@copyrightowner}{The Authors. This is the author's version of the work. It is posted here for your personal use. Not for redistribution. The definitive Version of Record will appear soon.}
\makeatother

\acmConference[QUASAR'25]{2nd Workshop on Quantum Algorithms, Software and Applied Research}{July 20 2025}{Notre Dame, IN, USA}

\title{High-level quantum algorithm programming using Silq}
\author{Viktorija Bezganovic}
\author{Marco Lewis}
\orcid{0000-0002-4893-7658}
\additionalaffiliation{
    \institution{Université Paris-Saclay, CNRS, CentraleSupélec, ENS Paris-Saclay, Inria, Laboratoire Méthodes Formelles}
    \city{Gif-sur-Yvette}
    \country{France}
    \postcode{91190}
}
\affiliation{
    \department{School of Computing}
    \institution{Newcastle University}
    \streetaddress{1 Science Square}
    \city{Newcastle upon Tyne}
    \country{UK}
    \postcode{NE4 5TG}
}
\author{Sadegh Soudjani}
\orcid{0000-0003-1922-6678}
\affiliation{
    \institution{Max Planck Institute for Software Systems}
    \streetaddress{Paul Ehrlich Str. 26}
    \city{Kaiserslautern}
    \country{Germany}
    \postcode{67663}
}
\author{Paolo Zuliani}
\orcid{0000-0001-6033-5919}
\affiliation{
    \department{Dipartimento di Informatica}
    \institution{Università di Roma ``La Sapienza''}
    \streetaddress{via Salaria 133}
    \city{Rome}
    \country{Italy}
    \postcode{00198}
}



\begin{abstract}
Quantum computing, with its vast potential, is fundamentally shaped by the intricacies of quantum mechanics, which both empower and constrain its capabilities. The development of a universal, robust quantum programming language has emerged as a key research focus in this rapidly evolving field. This paper explores Silq, a recent high-level quantum programming language, highlighting its strengths and unique features. We aim to share our insights on designing and implementing high-level quantum algorithms using Silq, demonstrating its practical applications and advantages for quantum programming.
\end{abstract}

\begin{document}

\maketitle

\section{Introduction}
\label{sec:intro}

Since its inception in the 1980s, quantum computing has grown into one of the most challenging yet promising areas of computer science, with the potential to revolutionize problem-solving and tackle complex computations far beyond the reach of classical computers. However, the process of developing software for quantum systems is fundamentally different from traditional computing, requiring new paradigms and approaches. Quantum programming languages, which are still in the early stages of development, face numerous challenges that must be resolved to fully leverage the unique computational advantages of quantum devices. As a result, advancing these languages is critical to unlocking the true potential of quantum computing.

Firstly, the quantum programming area suffers from a shortage of high-level abstractions. This results in great complications for developers while working with quantum software, as the developers are mostly forced to operate on quantum systems using low-level programming, which often gets overly complicated and inefficient. Expanding the range of available abstract high-level quantum languages would greatly benefit the quantum computing industry, as well as attract more software developers to the workforce. Moreover, abstractions in programming promote code reuse, which significantly speeds up the development process.

Secondly, faulty quantum computation can often be the result of erroneous software. Whether coming from the wrong algorithm implementation or bad outputs of quantum circuits, outcomes of bad computations hold a significant influence over the performance of quantum devices. To solve this problem, formal verification techniques and their implementation in quantum programs and systems are attracting the attention of researchers -- see for example the recent surveys \cite{Lewis23survey,chareton_handbook_formal_analysis}. However, verifiable programming languages are yet to be developed to fully utilize the advantages of quantum computation.

Finally, another significant issue of modern quantum programming is its limited resource management. Due to the novelty of quantum development, many limitations including the lack of resources (available qubits, memory, \etc) to support quantum computation remain unresolved. In order to operate on quantum hardware, strict resource management systems should be included in every developed program, which can be done by ensuring uncomputation and freeing up previously used resources.

This paper aims to introduce an alternative way of quantum software development using the Silq programming language while addressing several crucial issues within the current quantum development landscape. This is done by implementing a selection of non-trivial quantum algorithms and practically demonstrating the advantages of Silq.  

\section{Related Work}

Researchers have already addressed certain quantum mechanical aspects and complications of programming. One such example is Tower \cite{tower} - a quantum programming language that  supports data structures whose operations correspond to unitary operators in order to manipulate quantum superposition correctly. Contemporary quantum programming languages form abstractions for individual qubits and fundamental data types such as integers. In contrast, advanced quantum languages incorporate abstract data structures to speed up implementation and improve efficiency. In particular, Tower emphasizes the importance of supporting pointer-based, linked data featuring reversible semantics and allowing programs to be converted into unitary quantum circuits.

Entanglement is another quantum feature that needs to be taken into consideration. While critical to quantum computational advantage, handling entangled qubits requires additional verification steps. For example, various algorithms entangle qubits with temporary qubits that are eventually discarded, which might result in computational errors. To avoid such errors, the concept of state purity verification and assertion was introduced by the Twist programming language~\cite{Yuan_2022}. Twist introduces a type system that distinguishes expressions as a pure type, utilizes purity assertions to note the absence of entanglement, and employs a combination of static analysis and runtime verification to ensure the accuracy of purity specifications.

Recently, several quantum development tools have been widely recognized within the area of quantum programming. Languages such as Qiskit~\cite{9388498}, Cirq~\cite{isakov2021simulations}, Q\# \cite{Singhal_2023}, and Quipper \cite{Green_2013} have been proven to support a variety of quantum algorithms. However, there are still limitations that severely affect the development process. For instance, running Shor's algorithm~\cite{Shor97} on IBM Q Experience failed due to computational complexity and non-negligible noise. The algorithm's correctness is heavily based on its circuit design, so the solution would be to perform an in-depth theoretical analysis, as well as implement verification such as the methodology proposed in \cite{peng2022formally}.

Another complication of certain quantum programming languages is cluttered and unintuitive code. This issue often appears due to the necessity of creating additional helper functions, such as type casting or uncomputation functions required for quantum programs. The latter was addressed by the recent development of the high-level quantum programming language Silq~\cite{Bichsel2020} by implementing automatic uncomputation. While the focus of this paper is Silq and its practical implementation of quantum algorithms, recently published surveys \cite{Cartiere2022,q_programming_languages,qpl_survey_bibliography} provide detailed analyses of different quantum programming languages and are a strong recommendation for a reader choosing the tool for their project needs.

Table~\ref{tab:prog-lang} gives a summary of the different programming languages mentioned. To expand on some of the properties, a low abstraction level means that programs written in a language closer to describing circuits rather than the algorithm they represent, whereas it is the opposite for a high level of abstraction. For uncomputation, several programming languages have some means of doing automatic uncomputation, but this usually involves writing a certain expression within the program (see for example in Section~\ref{sec:silq:uncomp}), whereas automatic computation relies on types or annotations to variables to perform the uncomputation automatically. Finally, research-level programming languages tend to have a lower adoption rate (due to being unusable with quantum hardware), but explore new ideas for future programming languages that demonstrate some benefit. Those that are used by industry have ample access to simulators, hardware, IDE support, \etc


\begin{table*}[t]
    \centering
    \caption{A sample of quantum programming languages, their properties, and features.}
    \begin{tabular}{|c|p{1.5cm}|p{1.5cm}|c|c|p{4.5cm}|}
    \hline
    Programming Language & \centering Language Type & \centering Abstraction Level & Uncomputation & Usage & Unique Feature \\ \hline
    Silq~\cite{Bichsel2020} & Imperative & High & Automatic & Research & Type safety \par Automatic uncomputation \\ \hline
    Tower~\cite{tower} & Imperative & High & Partial & Research & Implementing data structures (sets, lists, etc.) as quantum types \\ \hline
    Twist~\cite{Yuan_2022} & Functional & Medium & Automatic & Research & Reasoning about purity embedded in type system \\ \hline
    Qiskit~\cite{qiskit2024}, Cirq~\cite{isakov2021simulations} & Python Framework & Low & Manual & Industry & Access to a variety of hardware\par Industry adoption \\ \hline
    Q\#~\cite{qsharp} & Imperative & High & Partial & Industry & Integrated in development kit\par Hybrid classical-quantum computing \\ \hline
    Quipper~\cite{Quipper} & Functional & Low (embedded in Haskell) & Manual & Research & Representation of complex quantum algorithms\par Dynamic lifting (use of measurement result in circuit) \\ \hline 
    \end{tabular}
    \label{tab:prog-lang}
\end{table*}

\section{Silq Programming}
\label{sec:silq}
In this Section, details of the Silq programming language~\cite{Bichsel2020} are described.
Silq was designed to address the challenge of unintuitive and cluttered low-level programming approaches by supporting safe and automatic uncomputation \cite{Bichsel2020}.

\subsection{Data Types}
\label{subsec:data-types}

A significant highlight of Silq as a programming language is its support of classical development approaches, which enables the usage of classical data types and the creation of hybrid quantum-classical programs. As demonstrated in Table~\ref{tab:data-types} below, certain data types are hybrid and can be utilized in both classical and quantum settings.

\begin{table}[t]
\centering
\caption{Silq data types~\cite{Bichsel2020}}
\label{tab:data-types}
\begin{tabular}{| p{.15\columnwidth} | p{.75\columnwidth} |}
 \hline
 Data type & Description  \\  
 \hline
 \hline
 1 & The singleton type that only contains element $()$.\\
 \hline
 $\mathbb{B}$ or B & Boolean values. It can be denoted classically as 1 or 0 (True or False), or non-classically as 1, 0 or any other state (superposition of states).\\
 \hline
 $\mathbb{N}$ or N & Natural numbers \{0, 1, \ldots\}. Can only be used classically ($!\mathbb{N}$).\\
 \hline
 $\mathbb{Z}$ or Z & Integer values \{\ldots -1, 0, 1, \ldots\}. Can only be used classically ($!\mathbb{Z}$).\\
 \hline
 $\mathbb{Q}$ or Q & Rational numbers, can only be used classically ($!\mathbb{Q}$).\\
 \hline
 $\mathbb{R}$ or R & Real numbers, can only be used classically ($!\mathbb{R}$).\\
 \hline
 \texttt{int[n]} & N-bit signed integers.\\
 \hline
 \texttt{uint[n]} & N-bit unsigned integers.\\
 \hline
 $\tau[]$ & Dynamic-length array.\\
 \hline
 $\tau^n$ & Vector of length n.\\
 \hline
 $\tau$×\ldots×$\tau$ & Tuple types, for example $!\mathbb{B}$×int[n], $!\mathbb{R}$×int[n].\\
 \hline
\end{tabular}
\end{table}

\subsection{Annotations}

Program annotations are an essential building tool for any software. Annotations are tags storing metadata on how certain structural program elements such as methods and variables should be handled. Due to the complex nature of quantum information, program annotations become a key component in ensuring proper data handling. 

Silq features several different annotations to classify data types and function behaviour.
\begin{description}
        \item[\textit{Classical types: \texttt{!}}]\hfill \\
        As mentioned in Section \ref{subsec:data-types}, Silq supports both classical and non-classical types of data. Classical types are specified using an exclamation mark before the variable type and imply the exclusion of the superposition of values. Variables are assumed to be of a quantum type by default. Even though it is not allowed to convert a quantum type to a classical type due to the presence of superpositions, classical types can be represented as quantum types by type casting. 

        \item[\textit{qfree}]\hfill \\
        The \texttt{qfree} function annotation indicates the function does neither introduce nor destroy superpositions. The important aspect of this annotation is the support for automatic uncomputation, which is a crucial aspect in the development of a quantum program.

        \item[\textit{mfree}]\hfill \\
        The \texttt{mfree} function annotation indicates the function can be executed without performing any quantum measurements.

        \item[\textit{const}]\hfill \\
        The variable annotation \texttt{const} indicates a variable that will not be changed through the execution process.

        \item[\textit{lifted}]\hfill \\
        The \texttt{lifted} expression indicates that the function is \texttt{qfree} (does not operate with superpositions) and the function's arguments are only constant. This annotation is essential for uncomputation. As it was previously mentioned, due to quantum mechanical properties the removal of temporary values within quantum code creates a threat of implicit measurement. Therefore, the \texttt{lifted} function expression makes arguments constant and enables automatic uncomputation by dropping temporary constants.
\end{description}

\subsection{Functions}
Due to a combination of classical and quantum programming in Silq, built-in functions supporting both types of computation are mandatory for the development process. The variety of classical functions includes mathematical (algebraic operations, exponentiation, comparators, \etc), logical and binary operators. In order to perform quantum computations, Silq supports the basic quantum operations (\texttt{H}, \texttt{X}, \texttt{rotX}, \ldots), measurement (through the \texttt{measure(q)} function), applying a phase to the quantum state (through \texttt{phase(r)}), and forgetting of a quantum variable if it is equal to some value (through \texttt{forget(x=y)}).

In addition to array and vector initialization, Silq also supports the creation of registers, consisting of multiple classical or quantum bits. Registers are initialized through the use of \texttt{int/uint/$\times$} data types. Similarly to arrays and vectors, registers are iterable code elements, which allow developers to easily access individual bits through the use of square brackets (int[n], uint[n]). This feature becomes particularly useful when the program needs to iterate through the binary representation of a decimal value (where the decimal value is initiated as a register).

\subsection{Safe Automatic Uncomputation}
\label{sec:silq:uncomp}
Automatic safe uncomputation in Silq is achieved through reverse reconstruction of temporary variables. The function in which the uncomputation needs to be performed must obey specific annotations (be of a \textbf{lifted} type) in order to allow safe automatic uncomputation.

Other programming languages have started to include means of handling uncomputation, but they can still be quite manual or restrictive. For instance, QSharp~\cite{qsharp} uses within-apply statements, \texttt{within \{\dots\} apply \{\dots\}}, to achieve uncomputation.\footnote{\url{https://learn.microsoft.com/en-us/azure/quantum/user-guide/language/expressions/conjugations}; accessed 13/03/2025.}
The statements that are provided in the \texttt{within} block are run, then the statements within the \texttt{apply} statement is run after, and finally the reverse of the statements in the \texttt{within} block are run to achieve uncomputation. This design has the drawback of needing to use multiple within-apply blocks when different states of the quantum variable to be uncomputed are needed. In Silq, the user can interleave statements between those that would need to be in a \texttt{within} or \texttt{apply} block, and the language will automatically detect how to uncompute it.

\section{Implemented Algorithms}

Due to Silq's technical strengths (support for both quantum and classical data types, improved handling of structural program elements, automatic uncomputation, error handling, \etc) and intuitiveness, Silq was chosen to develop this project and analyze its advantages in practice.

To demonstrate the capabilities of Silq, this section covers a selection of algorithms implemented using Silq; the code is available at~\url{https://github.com/v-bezganovic/silq-quantum-algorithms}.

\subsection{Unordered Quantum Minima Search}
\label{subsec:durr-hoyer}

One of the most common and frequently encountered problems while working with information storage includes element search in unsorted arrays. Considering an unordered list $T$ of length $N$ consisting of distinct integer values, the goal of the unordered quantum minima search algorithm is to determine an index $i$ such that $T[i]$ is a minimum value of the list $T$.

The classical searching approach heavily relies upon random selection and verification of values \cite{durr1999quantum}. As the required range of search expands (for example, as more records are added to a database), the number of queries required to perform the operations grows linearly, $O(N)$, losing its effectiveness.

The \DurrHoyer quantum search algorithm~\cite{durr1999quantum} (see Algorithm~\ref{alg:durr-hoyer}) requires $O(\sqrt{N})$ steps to analyze the unsorted list, using Grover's search~\cite{Grover96} as a subroutine.\footnote{Note that in Algorithm~\ref{alg:durr-hoyer}, the minima is returned rather than the index of the minima in the original algorithm. This can be easily changed by storing the measured index rather than the table value.}
Optimized search time is achieved by limiting the runtime and updating the oracle function with suitable solutions, detected using an amplitude amplification procedure (amplitude amplification is described in detail in Appendix~\ref{subsec:amplitudes}). The runtime is limited in that the runtime only increases when (1) a Hadamard operation is performed on a single qubit during the preparation stage, or (2) when a single step of the amplitude amplification operation is performed (\emph{i.e.}, applying the oracle and the diffusion operation). It can be shown that $\lceil 22.5\sqrt{N} + 1.4(\log_2 N)^2 \rceil$ steps suffice to determine the solution with high probability.

\begin{algorithm}[t!]
    \caption{\DurrHoyer Algorithm}
    \label{alg:durr-hoyer}
    \SetKwInOut{Input}{input}
    \Input{$T \: = \: [t_1, t_2, ... , t_N]$ \tcp{An unordered list $T$ of length $N$}}
    \tcp{Stage 0: initialization}
    $i = \text{rand}(1, N)$ \\
    $solution \: = \: T[i]$ \\
    Set $O_f$ such that $f(x) = T[x] \leq solution$ \\
    $n = \lceil \log_2{N} \rceil$, $q = \ket{0}^n$\\
    $stage = 0$, $rt = 0$ \\
    \While{$\text{rt } \leq \lceil \, 22.5\sqrt{N} + 1.4 (\log_2 N)^2 \rceil$}{
        \tcp{Stage 1: Prepare superposition}
        \If{$stage < n$}{$q[stage] = Hq[stage]$\\$stage += 1, rt+=1$}
        \tcp{Stage 2: Perform Grover search}
        \If{$n \leq stage < n + iterations$}{$q = D O_f q$ \\ $stage += 1, rt+=1$}
        \tcp{Stage 3: update result}
        \If{$stage == n+iterations$}{
        $y = \textbf{measure}(q)$ \\
        \If{$T[y] < solution$}{
            $solution = T[y]$ \\
            Set $O_f$ such that $f(x) = T[x] \leq T[y]$
        }
        $q = \ket{0}^n$, $stage = 0$
        }
    }
    $y = \textbf{measure}(q)$ \\
    \lIf{$T[y] < solution$}{$solution = T[y]$}
    \Return{$solution$}
\end{algorithm}

The algorithm's implementation was simplified by the use of Silq due to the support for a hybrid development approach. Since the algorithm operates on classical arrays using quantum principles and uses another quantum algorithm as a subroutine, the programming language of choice should support both classical and non-classical data types, which makes Silq a great fit for this algorithm.

\begin{figure}[t]
    \centering
\includegraphics[width=.85\columnwidth]{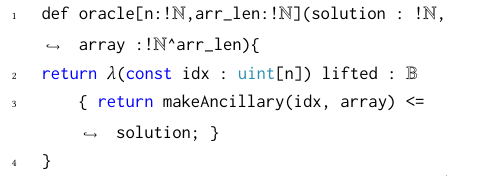}
    \caption{Oracle creation for Minima Search Algorithm}
    \label{fig:minima-oracle}
    \Description{Function for creating an oracle used for finding the minima. The function takes a solution (a positive integer) and array (of positive integers) as input and returns a function that checks if the value at an index in the array is less than the solution.}
\end{figure}

It must be noted, that the accuracy of Grover's search increases due to Silq's safe automatic uncomputation, proving its importance in the development process. Without the uncomputation, the drop of temporary variables required in Grover's search results in an implicit measurement that collapses the state.
This measurement, if done while the ancillary is entangled, could result in the amplitudes of the quantum state changing, affecting the likelihood of a marked state being measured  -- this highlights the importance of uncomputation.
Automatic safe uncomputation is possible by specifying the oracle used for the Grover search as \textit{lifted}, specifying it depends only on constant variables and utilizes functions that are \textit{qfree}, which neither introduce nor destroy superpositions. This can be seen in Figure~\ref{fig:minima-oracle}, where the created function is lifted and the ancillary made in the \texttt{makeAncillary} function is automatically uncomputed.

\subsubsection{Comparison to other languages}
The \DurrHoyer algorithm has previously been independently implemented, which helped to compare the algorithm implementation in Silq with Qiskit (see, for example, \cite{github}).

Firstly, after analyzing several projects available online, it should be noted that due to the recent Qiskit $2.0$ update, the code of older projects might require adaptation to new package requirements. While attempting to run the code, import statements and job execution required changing, as some built-in methods were no longer available.

Secondly, while execution is equally fast in both cases as both Silq and Qiskit are compiled languages, running the code on the Qiskit simulator required an additional transpilation step. Since Qiskit is a circuit-based language, in some cases circuits need to be transpiled before execution to fit the architecture of the desired backend. Additionally, the chosen backend has to be imported and specified before the execution. In contrast, Silq's simulator is installed alongside the coding extension and does not require additional setup.

Finally, to ensure the correct execution and prevent explicit measurements, Silq implementation utilizes the automatic uncomputation function by using \texttt{lifted} expression in Grover's subroutine. While Qiskit provides useful tools, such as the built-in Grover Operator class, the safe automatic uncomputation is yet to be addressed. 

\subsection{Collision Detection}
\label{subsec:collision}

The collision detection algorithm~\cite{Brassard_1998} is as follows: given an \textit{r-to-one} (or arbitrary) function $F \; : \; X \rightarrow \; Y$ that maps $r$ inputs to the same output, the aim of the algorithm (see Algorithm~\ref{alg:collision}) is to detect unique function inputs $x_0$ and $x_1$, such that $F(x_0) \; = \; F(x_1)$. The classical approach would require exhaustive checking of all the inputs, resulting in a long runtime. The quantum approach offers an efficient alternative with a high success probability, which requires only $O(\sqrt[3]{N/r})$ function evaluations.

\newcommand{\lIfElse}[3]{\lIf{#1}{#2 \textbf{else}~#3}}
\DontPrintSemicolon
\begin{algorithm}[t!]
    \caption{Quantum Algorithm for the Collision Problem}
    \label{alg:collision}
    \SetKwInOut{Input}{input}
    \Input{$T \: = \: [t_1, t_2, ... , t_N], F : T \to Y, r \in \mathbb{N}$}
    \tcp{Initialization}
        \lIfElse{$r \geq 2$}{$k = \sqrt[3]{N/r}$}{$k = rand(1,N)$}
        Generate a subset $S \subseteq T$ using \textbf{generateSubset} function with cardinality $k$\\
        Generate lists \emph{input} from $T$ and \emph{output} from $Y$ using \textbf{generateLists} function \\
        collision1 : $!\mathbb{N}$, collision2 : $!\mathbb{N}$ \\
    \tcc{Initial observation and Grover's search}
    Check for a collision in \emph{output} \\
    \eIf{no collision}{
        Create $H : N \to \{0,1\}$ such that $H(x) = 1$ if there exists $input[i] \neq T[x]$ such that $output[i] = F(T[x])$ and $H(x) = 0$ otherwise \\
        \lIfElse{$r \geq 2$}{$t = (r-1) * k$}{$t = 1$}
        pendingSolutionIndex = \textbf{grover($H$, $t$)}\\
        \For{i \textbf{in} $[0..k)$}{
            \If{$F(T[pendingSolutionIndex]) == outputs[i]$ \textbf{and} $T[pendingSolutionIndex] != inputs[i]$}
            {$collision1 = T[pendingSolutionIndex]$\\
            $collision2 = inputs[i]$}
        }
    }
    {
        Set $collision1, collision2$ to the indexes of the colliding entries
    }
    \Return{(collision1,collision2)}
\end{algorithm}

The initial step is to prepare the input data for analysis. An array of natural numbers $T$ is used to represent the set $X$ that is going to be searched for collisions.
The oracle $F$ is up to the user to specify.
For instance, if the function $F$ computes $x \; mod \; 5$ then the algorithm will return the colliding elements that result in the same output after calculating $x \; mod \; 5$. The variable $r$ is used to represent that $F$ is {$r$-to-one} and is also given by the user (if $F$ is arbitrary, then $r$ must be set to $0$ or $1$).

To begin the procedure, a random subset of numbers $S$ of cardinality $k$ is taken from $T$ using the function \texttt{generateSubset}. The subset may contain inputs that collide on $F$; if the duplicates are present within the initial subset, the algorithm will detect them (as we will see).
Since the algorithm determines the elements based on the function's output, an array of outputs $Y'$ is generated based on the subset $S$, calculated as
\begin{equation*}
\begin{aligned}
        & F(S)\: = \: Y' \subseteq Y, \\
        & S \: = \: [s_1, s_2, \ldots, s_k], \\
        & Y' \: = \: [F(s_1), F(s_2), \ldots, F(s_k)].
\end{aligned}
\label{eq:set-gen}
\end{equation*}
In the code, the generation is done with the function \texttt{generateLists}, returning input and output lists with respectively stored values. Finally, variables $collision1$ and $collision2$ are created as result placeholders for colliding elements.

\begin{figure}[t]
    \centering
    \includegraphics[width=.85\columnwidth]{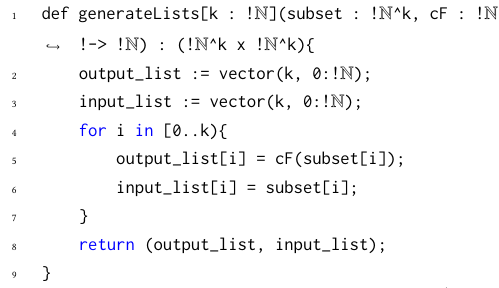}
    \caption{\texttt{generateLists()} function for the Collision Detection Algorithm}
    \label{fig:generate-collision}
    \Description{Function that generates two lists. One consists of the input values from the subset. The other list consists of the evaluation of those values on the function to be checked for collisions.}
\end{figure}

The \texttt{generateList} (Figure~\ref{fig:generate-collision}) function takes the subset as an input with its cardinality $k$ and computes input and output lists for later analysis using the provided oracle. A classical variant of the oracle $F$, denoted $\texttt{cF}$, is used to correctly handle types in the classical variant of the code; this oracle is automatically generated within the implementation and calls $F$ whilst casting between types. Initially, two lists $input\_list$ and $output\_list$ are generated to store the elements. Then, the iteration through the subset is performed, taking values from the subset directly to the input list and oracle values to the output list.

\begin{figure}[t]
    \centering
    \includegraphics[width=.85\columnwidth]{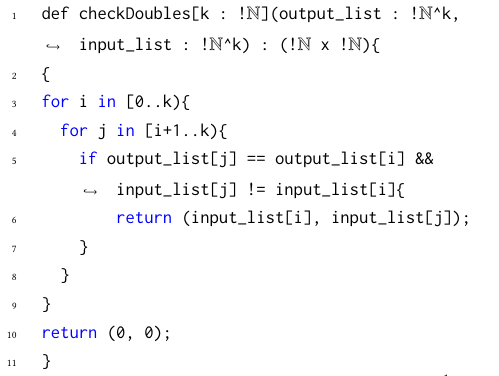}
    \caption{\texttt{checkDoubles()} function for the Collision Detection Algorithm}
    \label{fig:doubles-collision}
    \Description{A function that checks if two entries in the output list are the same and the entries of the input list are different.}
\end{figure}

The function \texttt{checkDoubles} (Figure~\ref{fig:doubles-collision}) takes the previously created input/output sets, and checks for collisions within the subset. Since the subset's cardinality is much smaller than the original set's, the classical method of sequential iteration through the set can be used instead of the quantum approach. As mentioned before, after the setup the observations using the \texttt{checkDoubles} method must be performed on the generated subset to verify that the initial subset does not contain any duplicates. This is done to determine whether any elements within the subset match the collision criterion. If colliding elements are detected, the program terminates early and returns such elements.

\begin{figure}[t]
    \centering
    \includegraphics[width=.85\columnwidth]{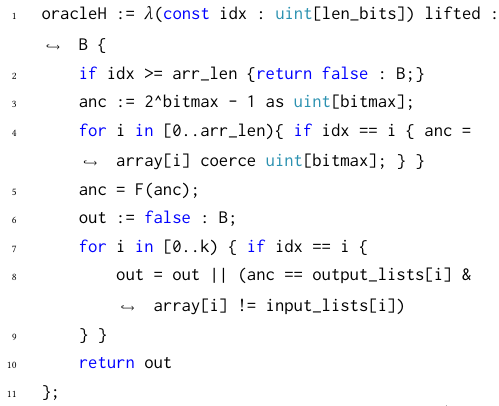}
    \caption{Oracle for Collision Detection Algorithm}
    \label{fig:oracle-collision}
    \Description{The oracle used for the Grover's part of the collision detection algorithm. It returns 1 whenever there is a collision at an index with any of the other values not in the subset.}
\end{figure}

In case of no initial detection, further calculations are performed. Since the collision detection algorithm utilizes Grover's search as a subroutine. A new oracle $H$ (Figure~\ref{fig:oracle-collision}, $H$ denoted \texttt{oracleH}) is generated based on the oracle $F$ that searches over the indexes of $T$. The oracle $H$ stores the values of $T$ in an ancillary quantum register and then performs the oracle $F$ on that ancillary register. From there, the value of the ancillary register is compared against the values in the generated subset.
For the collision detection algorithm, the oracle function must check whether the index satisfies the following conditions:
\begin{enumerate}
    \item $F(T[x]) \: \in \: Y'$,
    \item $T[x] \notin S$.
\end{enumerate}
Additionally, it should be noted that the oracle makes use of Silq's uncomputation capabilities to automatically uncompute the ancillary register that was used to represent the value.

Grover's search is performed to determine the index of the solution. The implementation used differs slightly from the Silq implementation (requiring an additional input) as the number of iterations can be reduced by providing the number of marks, which can be determined based on $r$ and $k$. By running \texttt{grover($H$, $t$)}, the temporary solution's index is determined and it can be verified classically through conditional checking.

Because the algorithm heavily relies on the random generation of subsets to perform collision checks, the abstract high-level approach of Silq and its support for hybrid data types enables convenient and quick ways to perform randomization. The randomization function can be defined similarly to the classical programming approach and easily integrated into other algorithms. In contrast, the circuit-based approach using languages similar to Qiskit would require defining a randomization procedure as a sub-circuit and incorporating it into the main circuit, which would be more cumbersome to implement. Moreover, Silq's hybrid programming approach enables flexible work with both classical and quantum data types.

Similarly to the \DurrHoyer algorithm, the Collision Detection algorithm performs the selection of comparison elements using Grover's search subroutine. As previously described in Section \ref{subsec:durr-hoyer}, Silq's automatic uncomputation increases the accuracy of the algorithm by preventing undesired implicit measurement.

\subsection{Uniform Superposition}
\label{subsec:uniformsup}
Uniform superposition preparation is a crucial initial component of many quantum algorithms, including Grover's search~\cite{Grover96} or Shor's factoring algorithm~\cite{Shor97}.

\begin{figure}
\centering
    \includegraphics[width=.85\columnwidth]{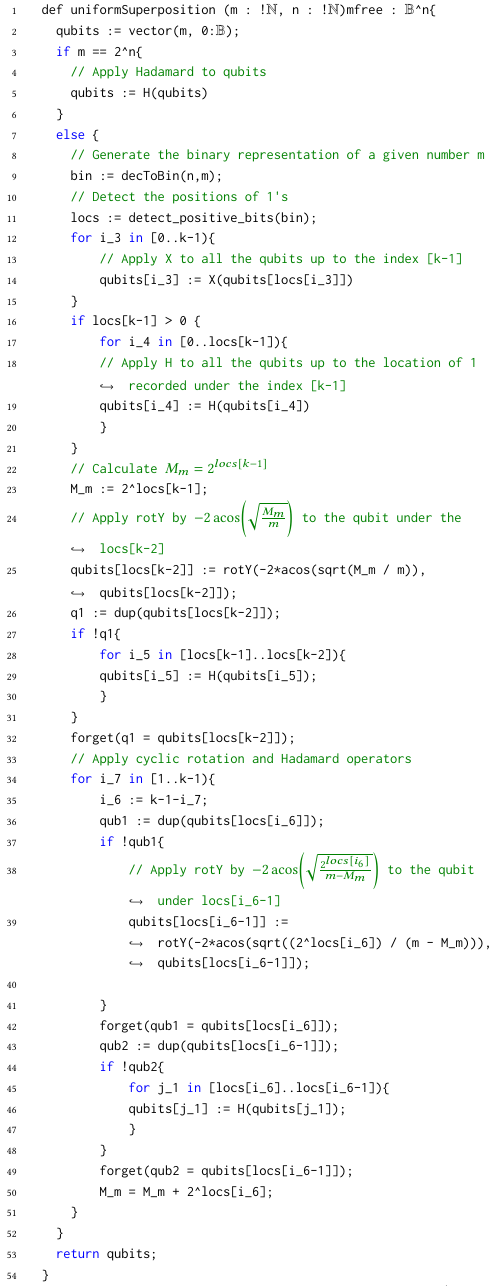}
    \caption{Uniform Superposition Preparation Algorithm}
    \label{alg:superposition}
    \Description{The uniform superposition preparation algorithm written in Silq, which given a positive integer, generates a uniform superposition of all positive integers up to the one provided.}
\end{figure}

While the standard method of uniform superposition state preparation using the Hadamard operator is optimal for $M=2^{n}$ states, a different approach needs to be taken when $M \neq 2^{n}$ states.
The uniform superposition state preparation algorithm~\cite{shukla2023efficient} (see Figure~\ref{alg:superposition}) introduces an alternative procedure to obtain the required superposition state $|\psi\rangle = \frac{1}{\sqrt{M}} \sum_{j=0} ^{M-1} |j\rangle$, where $M$ represents the number of distinct states within the superposition state given that $2 < M < 2^n$.
As far as we are aware, there are no other implementations of this algorithm currently.

The algorithm's input is a positive integer $M$, such as $M \neq 2^n$ for any $n \in \mathbb{N}$. The algorithm starts with the generation of the binary representation of $M$. It must be noted, that due to operations on the binary expression $M_{bin}$, the calculations are performed in the reverse order, starting with the least significant bit. 

After generating the binary representation of $M$, the locations of positive bits (\emph{i.e.}, with value 1) are recorded.
The next step is initializing the array of qubits, the cardinality of which is determined based on the user's input and is equal to $\lceil \log_2{M} \rceil$  -- see  \eqref{eq:qubits}: 
\begin{equation}
\label{eq:qubits}
    Q = [q_1, q_2, \ldots, q_k], \quad k = \lceil \log_2{M} \rceil .
\end{equation}

Using the previously generated positive bit locations, the \texttt{X} (NOT) operator is applied to qubits under respective indices. This is done to encode the user's input into the generated quantum state to perform quantum operations. 

The main operators utilized in the rest of the algorithm are the Hadamard, the controlled Hadamard and the rotation around the Y-axis \eqref{eq:roty}:
\def\rot{
\begin{bmatrix}
cos(\frac{\theta}{2}) & -sin(\frac{\theta}{2})\\
sin(\frac{\theta}{2}) & cos(\frac{\theta}{2})  
\end{bmatrix} . 
}

\begin{equation}
\label{eq:roty}
    R_Y(\theta) = \rot 
\end{equation}

Our implementation of the algorithm demonstrates the important features of Silq practically, such as variable uncomputation using the \texttt{forget} statement (see Figure~\ref{alg:superposition}, lines 32, 42, and 49). During the conditional rotation and Hadamard application phase, the algorithm takes qubit values to determine the action to be performed. Due to the inability to reuse quantum variables, the developer is required to create a duplicate of an existent quantum variable to use it for conditional checking, which would result in a computational resource shortage. To avoid that, the built-in \texttt{forget} function can be used immediately after the conditional checking, which uncomputes the variable and frees up resources to be used in later computations. However, the safety of this uncomputation is not guaranteed, therefore it is up to the developer to ensure its correct execution.

\section{Discussion}

The demonstrated algorithm implementations highlighted the variety of advantages of Silq. First and foremost, the support for both classical and quantum variable types enables hybrid programming, which is a great benefit for algorithm development. Additionally, easy type-casting removes the majority of complications while simultaneously working with both quantum and classical variables.

Secondly, automatic uncomputation of temporary variables provides a safe and efficient approach to managing the stability of the quantum state with less input from the developer. Automatic uncomputation was used in both the \DurrHoyer and the Collision Detection algorithm within Grover's subroutine to measure temporary value after performing Grover's diffusion. As a result, the retrieved value was safe from explicit measurement error and is therefore more reliable to use in further calculations. In addition, Silq facilitates a variety of additional functions, such as the \texttt{forget} function and the \texttt{dup} tool, which copies quantum variables without breaking the no-cloning theorem.

Although a quantitative analysis has not yet been performed, Silq has already been shown to reduce the number of line-code compared to Q\#~\cite{Bichsel2020}. A full analysis would require implementations across multiple languages for increasingly complex quantum programs. Previously, an analysis has been done with the standard quantum algorithms (Deutsch-Jozsa algorithm, Grover's algorithm, \etc) on several languages~\cite{expressiveness}. In the future, Silq could be compared to Quipper~\cite{Quipper} by implementing the Triangle Finding algorithm, covered in the paper.

Currently, Silq has several limitations. First and foremost is the general lack of libraries to perform some basic operations. For instance, while working on the implementation of the Collision Detection algorithm, random generations of numbers and subsets were performed manually with binary randomization using Hadamard gates. However, the helper function to perform randomization can be imported and re-used across programs. It must be noted that the function utilizes registers, described in Section \ref{sec:silq}. The limit of this function is integers under 30, but it can be scaled by extending the number of bits to fit the needs of the code.

Silq would further benefit from type features in other languages. This includes abstract types. For instance, the oracle function for collision detection, $F$, is required to have a type of $\texttt{uint[n]} \to \texttt{uint[n]}$ in the argument of the program. It would be more appropriate if $F$ could be of type $A \to B$ where $A$ and $B$ are generic quantum data types that could be used in the program. Currently, Silq's quantum data types are restricted to signed and unsigned integers, booleans, and tuples/vectors of those. Whilst in Silq this could be achieved by specifying the number of qubits to use for \texttt{uint} and \texttt{int} types (\emph{e.g.}, $\texttt{uint[a]} \to \texttt{uint[b]}$), future quantum programming languages may use a greater variety of types and, therefore, more general specification of types would be beneficial.

\section{Conclusion}

In this paper, we have introduced the features of Silq as a programming language and demonstrated its practical use by implementing the Quantum Minima Search, Collision Detection and Uniform Superposition Preparation algorithms.

Silq has proven to be an efficient tool for the development and testing of quantum algorithms. Despite its contrast with common circuit-oriented quantum programming frameworks such as Qiskit and Cirq, Silq provides efficient tools to manipulate quantum states. Additionally, the approach and syntax proposed by Silq are intuitive and easy to use for software developers.

Another great proven advantage of Silq is the variety of data types available to develop algorithms. Supporting both classical and quantum data types enabled hybrid development, which is crucial for several algorithms involving classical computations and post-processing. However, while having the benefits of hybrid development, it is crucial for the programmer to ensure proper data type handling and conversion.

Finally, the support for automatic uncomputation removes the need to manually add helper functions to perform uncomputation, which optimizes the development process and ensures the correctness of the performed calculations. Moreover, the absence of redundant helper methods results in less cluttered code, making the development process easier for programmers.

In the future, further research into high-level quantum algorithms is planned to evaluate the performance of Silq. Moreover, future research will cover other quantum programming languages to investigate the current state of the high-level quantum development landscape on a wider scale and to practically evaluate and compare the capabilities of available development tools.

\begin{acks}
The work of M. Lewis has been partially funded by the French National Research Agency (ANR) within the framework of ``Plan France 2030'', under the research projects EPIQ ANR-22-PETQ-0007, HQI-Acquisition ANR-22-PNCQ-0001 and HQI-R\&D ANR-22-PNCQ-0002; and partially by the UK Engineering and Physical Sciences Research Council (EPSRC project reference EP/T517914/1).
The work of S. Soudjani  was supported by the following grants: EIC 101070802 and ERC 101089047.
P. Zuliani was supported by the SERICS project (PE00000014) under the Italian MUR National Recovery and Resilience Plan funded by the European Union - NextGenerationEU. 
\end{acks}

\bibliographystyle{ACM-Reference-Format}
\bibliography{References/refs}

\appendix
\section{Background}

This section introduces the main concepts behind the development of the algorithms, as well as the mathematical notation used in the paper. For a full background, see for example~\cite{nielsen}.

\subsection{Notation}
\subsubsection{Quantum Bit} \hfill \\
A qubit, also known as a quantum bit, is the computational unit of quantum information used to supply information and communicate through the system. Unlike the classical bit which can be in states of either 1 or 0, a qubit can also be in states between 1 and 0.
The state of the qubit (also referred to as the simplest quantum state) is observed after performing measurement and can be described as 
\begin{equation*}
    |\psi\rangle = \alpha|0\rangle + \beta|1\rangle
\end{equation*}
where $\alpha, \beta \in \mathbb{C}$, $|\alpha|^2$ is the probability of obtaining the qubit state $|0\rangle$ and $|\beta|^2$ is the probability of obtaining qubit state $|1\rangle$.

\subsubsection{Multi-qubit Systems} \hfill \\
The state of a quantum system is described through normalized vectors in a complex vector space. Using the definition of a single-qubit system, the combined system of multiple qubits can be easily constructed and described using the tensor product notation. For example, a system of three qubits where the states of the individual qubits are $|\psi_{i}\rangle = \alpha_{i}|0\rangle + \beta_{i}|1\rangle$ for $i=1,2,3$, is described as
\begin{equation*}
\begin{aligned}
    & |\psi_1\psi_2\psi_3\rangle = |\psi_1\rangle \otimes |\psi_2\rangle \otimes |\psi_3\rangle
    \\ & = \alpha_1\alpha_2\alpha_3|000\rangle + \alpha_1\alpha_2\beta_3|001\rangle +
    \alpha_1\beta_2\alpha_3|010\rangle + \\ & \quad \alpha_1\beta_2\beta_3|011\rangle +
    \beta_1\alpha_2\alpha_3|100\rangle + \beta_1\alpha_2\beta_3|101\rangle + \\ & \quad
    \beta_1\beta_2\alpha_3|110\rangle + \beta_1\beta_2\beta_3|111\rangle.
\end{aligned}
\end{equation*}
\subsubsection{Quantum Operators}\hfill \\ 
Quantum programs are developed through manipulations of quantum states using specific operators, which is also referred to as quantum evolution. Operators can be expressed as matrices, which allows the description of quantum evolution using linear algebra.

For example, some of the operators used in developing algorithms described in this paper:

\def\X{
\begin{bmatrix}
    0 & 1 \\
    1 & 0
\end{bmatrix}}

\def\HH{
\begin{bmatrix}
    1 & 1 \\
    1 & -1
\end{bmatrix}}

\def\Y{
\begin{bmatrix}
    0 & -i \\
    i & 0
\end{bmatrix}}

\begin{equation}
    \begin{aligned}
    & \text{Hadamard}\; (H) = \frac{1}{\sqrt{2}}\HH, & \text{NOT}\; (X) = \X, \\
    & \sigma_y=\Y.
    \end{aligned}
\end{equation}

\subsection{Amplitude Amplification}
\label{subsec:amplitudes}

Amplitude amplification~\cite{Brassard02} is a quantum algorithm to distinguish the solution of a search problem by increasing its probability amplitude. The steps of the algorithm are described as follows:
\begin{enumerate}
    \item The initial step is the creation of a uniform superposition shown in Eq.~\eqref{eq:superposition} by applying the Hadamard operation on the whole range of states. As a result, all states obtain the same probability amplitude: 
    \begin{equation}
    \label{eq:superposition}
        |\psi\rangle \: = \: H^{\otimes n}|0\rangle^n .
    \end{equation}
    
    \item The next step is determining the candidate solution. Commonly, search algorithms use oracles to distinguish suitable entries. After the candidate solution is found, its probability amplitude is reversed and set to the negative value as shown in Eq.~\eqref{eq:reverse}. Consequently, the average amplitude of values is lowered.
    \begin{equation}
    \label{eq:reverse}
        U_F = I - 2P_f, \: P_f = \sum_{f(x)=1} \ket{x}\bra{x} .
    \end{equation}
    \item The final transformation step is applying an additional reflection $U_\psi$ on an amplitude of a candidate solution as follows: $U_\psi \: = \: 2|\psi\rangle\langle \psi|\:-\:I$. Since the main amplitude is previously lowered, the state $|\psi\rangle$ becomes closer to the candidate solution.        
\end{enumerate}

\section{Silq Functions}

\begin{table}[h]
\centering
\caption{Summary of quantum functions available in Silq (based on \url{https://silq.ethz.ch/documentation})}
\label{tab:functions}
\begin{tabular}{| p{.23\columnwidth} | p{.65\columnwidth} |}
 \hline
 Function & Description  \\  
 \hline
 \hline
 measure & Measure the state and return 0 or 1.~\\
 \hline
 H & Hadamard operator.
 \textit{H()} performs Hadamard transformation on a qubit.~\\
 \hline
 phase & \textit{phase(r)} rotates part of the quantum state by $r$ radians (multiplies the quantum state by $e^{ir}= \cos(r) + i \sin(r)$).\\
 \hline
 rotX, rotY, rotZ & The rotation operator that rotates a given state around the X, Y and Z axes respectively.~\\
 \hline
 X, Y, Z & The operator applies X, Y and Z gates to the state respectively.  \\
 \hline
 dup & The function duplicates the quantum state $|q\rangle \to |q\rangle|q\rangle$ without violating the no-cloning theorem.~\\
 \hline
 array, vector & Similarly to the duplication function, the functions \textit{array(m,v)} or \textit{vector(m,v)} return an array or a vector filled with \textit{m} duplicates of \textit{v}.\\
 \hline
 forget & The function can be both conditional and unconditional. The conditional \textit{forget(x,y)} forgets \textit{x} if it is equal to \textit{y}. Alternatively, \textit{forget(x)} attempts to unconditionally uncompute \textit{x}.\\
 \hline
\end{tabular}
\end{table}

\newpage

\section{Additional Code}

\begin{figure}[h]
    \centering
    \includegraphics[width=.85\columnwidth]{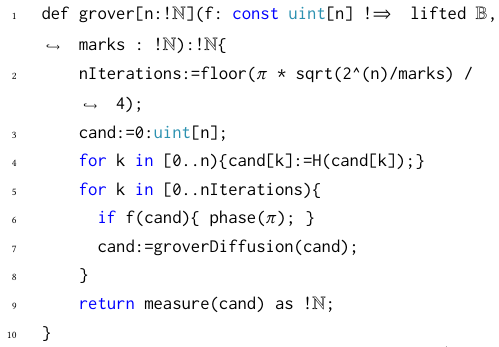}
    \caption{Modified Grover's Subroutine Algorithm~\cite{Bichsel2020}}
    \label{fig:grover}
    \Description{Grover's algorithm modified to allow for multiple marked elements.}
\end{figure}

\begin{figure}[h]
    \centering
    \includegraphics[width=.85\columnwidth]{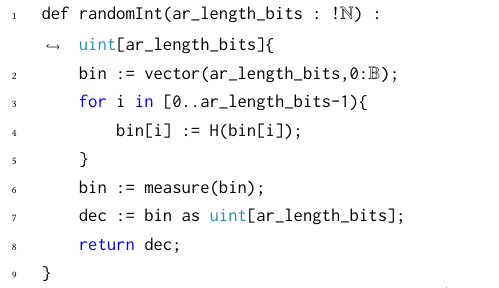}
    \caption{Integer randomization function using quantum superposition}
    \label{fig:int-random}
    \Description{A function for returning a random integer using Hadamard gates.}
\end{figure}

\begin{figure}[h]
    \centering
\includegraphics[width=.85\columnwidth]{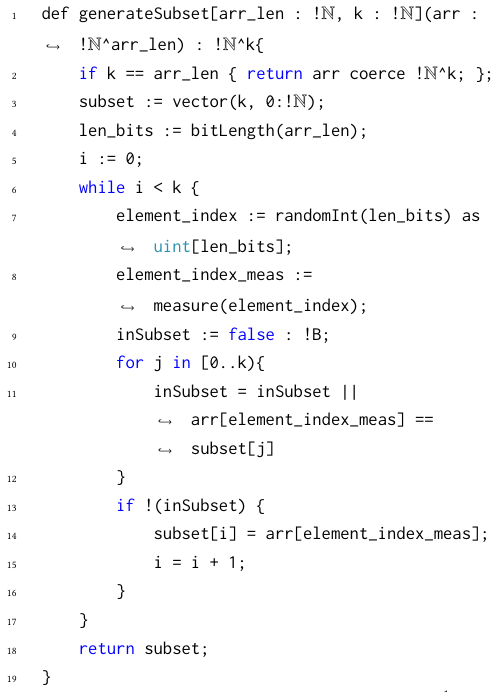}
\caption{\texttt{generateSubset()} function for the Collision Detection Algorithm}
\label{fig:subset-collision}
\Description{A function to generate a random subset of fixed length from an array.}
\end{figure}

Figures~\ref{fig:grover}, \ref{fig:int-random}, and \ref{fig:subset-collision} are functions used in the Collision Detection algorithm (Section~\ref{subsec:collision}). Figure~\ref{fig:grover} is the implementation of Grover's algorithm used. It is modified to account for how many marked values are in the oracle function. Figure~\ref{fig:int-random} implements integer randomization. Figure~\ref{fig:subset-collision} is the code used for generating a subset. 

Algorithm~\ref{alg:uniform-superposition} is the pseudo-code for the Uniform Superposition algorithm (Section~\ref{subsec:uniformsup}).

\newpage

\newpage

\begin{algorithm}[h]
    \caption{Uniform Superposition Algorithm}
    \label{alg:uniform-superposition}
        \textbf{Input}
        $M : \: !\mathbb{N}, \; 2 \leq M \leq 2^n$ \\
        \tcc{Uniform Superposition Algorithm}
        $Q \: : \: \mathbb{B}[n], \; Q=[q_1,q_2, \ldots, q_n]$ \tcp{Initialize the starting state $|\phi\rangle = \sum_{n=0} ^{M-1} |0\rangle$}
        $M_{bin} = [m_1, m_2, \ldots , m_n], \; m_i \in \mathbb{B}$ \tcp{Generate binary representation of $M$}
        \tcp{Detect locations of positive bits within the binary representation}
        locs : !$\mathbb{Z}$ \\
        currentSlot = n-1 \\
        loopIndex = 0 \\
        \While{currentSlot $>$ 0}{
        \If{the binary bit of $M_{bit}$ under \textit{currentSlot} is positive}{
        Record currentSlot in locs[loopIndex]\\
        Increase loopIndex}
        Reduce currentSlot}
        k = locs.length\\
        \tcp{Apply Hadamard to qubits under indexes stored in \textit{locs}}
        \If{locs[k-1] $>$ 0}{Apply Hadamard to qubits before locs[k-1]}
        Initialize $M_m \: = \:2^{locs[k-1]}$\\
        Apply \texttt{rotY($\theta$)} to the qubit $|q_{locs[k-2]}\rangle$ for $\theta = -2 \arccos(\sqrt{\frac{M_0}{M}})$\\
        \If{!$|q_{locs[k-2]}\rangle$}{Apply Hadamard to qubits between locs[k-1] and locs[k-2]}
        \tcc{Cyclic rotation and Hadamard application}
        \For{i in [1..k-1)}{j = k-1-i \\
        \If{!$|q_{locs[j]}\rangle$}{Apply \texttt{rotY($\theta$)} to qubit $|q_{locs[j-1]}\rangle$, for $\theta = -2\arccos(\sqrt{\frac{2^{locs[j]}}{M-M_m}})$}
        \If{!$|q_{locs[j]}\rangle$}{Apply Hadamard to qubits between locs[j] and locs[j-1]}
        $M_m = M_m + 2^{locs[j]}$}
        \Return{$Q$}
\end{algorithm}

\end{document}